\documentclass[a4paper,11pt]{article}
\pdfoutput=1 

\usepackage{jcappub} 

\usepackage[T1]{fontenc} 

\title{\boldmath Modified Compton Effect and CMB Anisotropy}

\author[a,b,1]{S. Davood Sadatian,\note{Corresponding author.}}
\author[b]{Amir Sabouri, }
\author[c]{Zahra Davari}

\affiliation[a]{Department of Physics, Faculty of Basic Sciences, University of Neyshabur , P. O. Box 9319774446, Neyshabur, Iran}
\affiliation[b]{Research Department of astronomy $\&$ cosmology, University of Neyshabur, P. O. Box 9319774446, Neyshabur, Iran}
\affiliation[c]{School of Physics, Korea Institute for Advanced Study (KIAS), 85 Hoegiro, Dongdaemun-gu, Seoul, 02455, Republic of Korea}

\emailAdd{sd-sadatian@um.ac.ir, sd-sadatian@neyshabur.ac.ir,asaburi@neyshabur.ac.ir,zahradavari@kias.re.kr}

\abstract{Recent satellite observations have revealed significant
anisotropy  in the cosmic microwave background (CMB) radiation, a
phenomenon that had previously been detected but received limited
attention due to its subtlety. With the advent of more precise
measurements from satellites, the extent of this anisotropy has
become increasingly apparent. This paper examines the CMB radiation
by reviewing past research on the causes of CMB anisotropy and
presents a new model to explain the observed temperature anisotropy
and the anisotropy in the correlation function between temperature
and E-mode polarization in the CMB radiation. The proposed model is
based on a modified-generalized Compton scattering approach
incorporating Loop Quantum Gravity (LQG). We begin by describing the
generalized Compton scattering and then discuss the CMB radiation in
the context of processes occurring at the last scattering surface.
Our findings are derived from the latest observational data from the
Planck satellite (2018). In our model, besides the parameters
available in the Planck data for the standard model ($\Lambda$CDM),
we introduce two novel parameters: $\delta_{L}$, the density of
cosmic electrons, and $M^2$, a parameter related to the
modified-generalized Compton scattering effects. The results
indicate that, based on the 2018 Planck data, small values were
obtained for $\delta_{L}$ and $M^{2}$,
$\delta_{L}=1.63\pm0.08(10^{-13})$ and $M^2=2.28\pm0.34(10^{-4})$,
showing no significant deviation from the standard model. Moreover,
increasing the values of $\delta_{L}$ and $M^{2}$ leads to an
increase in the range of fluctuations in the CMB temperature
anisotropy power spectrum and the correlation function between
temperature and E-mode polarization for multipoles $l<500$ until the
first peak.
\\\\
\texttt{PACS:}\,  04.50.kd ; 95.36.+x\\
\texttt{Key Words:}  Loop Quantum Gravity, Compton Effect, Cosmic Microwave
Background.}

\begin{document}
\maketitle
\flushbottom

\section{Introduction}
\label{sec:intro}
Loop Quantum Gravity (LQG) is a theoretical framework that aims to reconcile quantum mechanics and general relativity in a way that is consistent with the standard cosmological model.
In LQG, the structure of space-time is composed of entangled loop networks known as spin networks at scales larger than the Planck length ($l_{p}\sim10^{-35}$ m).
Due to the inconsistency of experimental data with the quantum aspects of gravity and the lack of a satisfactory candidate, the problem of general relativity merge and quantum mechanics still has major challenges \cite{1}.
In 1916, Einstein laid the groundwork for the concept of "quantum Riemannian geometry" to describe the quantization of gravity, which replaces the 4D space-time metric with a probability domain for various space-time geometries \cite{2}. The systematic development of LQG based on quantum Riemannian geometry was completed in the 1990s \cite{3,4,5,6,7}, addressing some critical issues in quantum gravity \cite{8,9}.\\
In the context of LQG cosmology, the structure of space-time does
not end at the Big Bang but extends to wider dimensions in the
quantum geometry of space-time \cite{10}. One of the promising
aspects of quantum gravity is the "asymptotic safety program"
proposed within the standard model of particle physics \cite{11}.
Since the inception of LQG research by pioneers such as Ashtekar,
Rovelli, Smolin, Bergmann, Dirac and Wheeler \cite{1,3,4,5,9,12,13},
and others, the theory has faced persistent challenges over the past
two decades, including:
\begin{itemize}
    \item How to quantize finite Hamiltonian systems without using background fields or perturbation techniques?
    \item What dynamic effects arise from the absence of a background space-time metric?
    \item How to precisely calculate quantum transition amplitudes, and how can quantum gravity naturally resolve curvature singularities of classical general relativity?
    \item Are the UV divergences of quantum field theory (QFT) cured\cite{1}?
\end{itemize}
Another topic of interest in LQG is the "Compton scattering effect". Compton scattering, a fundamental electromagnetic radiation phenomenon, occurs when a high-energy photon collides with an electron, resulting in a decrease in the photon's energy (frequency) and an increase in its wavelength, a phenomenon known as the "Compton shift" \cite{14}. To describe Compton scattering, light is modeled as a stream of particles (photons) with varying energies or frequencies \cite{15}. Given the low precision of Lorentz symmetry observed in nature, the violation of Lorentz invariance has been examined from multiple perspectives \cite{16,17,18,19,20,21,22,23,53}. In LQG, the Compton effect can be reformulated based on modified standard dispersion relations, reflecting Lorentz invariance violations in quantum gravity \cite{21,22}. However, these modified dispersion relations are not mandated by LQG theory itself but are derived heuristically, contingent on solving the problem of quantum gravity and matter dynamics \cite{23}.\\
 One of the most valuable applications of the LQG approach is determining the anisotropy of the cosmic microwave background radiation (CMB), which offers cosmologists crucial insights into the early universe's structure and evolution. The first observational data from the COBE satellite revealed anisotropies in CMB photons, especially in terms of temperature and polarization \cite{24}.  Subsequent satellites, such as WMAP and Planck, have provided more accurate measurements of CMB anisotropy \cite{25,26,27,28,29}.\\
 In this paper, we first review the modified dispersion relations and generalized Compton scattering effects based on LQG in section \ref{Sec2}. Section \ref{Sec3} discusses CMB anisotropy stemming from processes at the last scattering surface. In section \ref{Sec4}, we present a model incorporating generalized Compton scattering within the LQG framework to improve the accuracy of temperature variation calculations and the correlation function between temperature and E-mode polarization anisotropy of the CMB. Section \ref{Sec5} details the results derived from this model using the latest observational data from Planck \cite{28,29}. Finally, section \ref{Sec6} concludes the paper.
\section{Generalized Compton Scattering (GCS)}
\label{Sec2} In this section, we review the modified dispersion
relations of the generalized Compton effect from the LQG approach.
According to the conservation of linear momentum in Compton
scattering ($\vec{p}=\vec{\acute{p}}+\vec{P}$), where $\vec{p}$ and
$\vec{\acute{p}}$ are the linear momentum of the photon before and
after the collision with an electron at the rest state,
respectively, and $\vec{P}$ is the linear momentum of the electron
after the collision with the photon, it can be written:
 \begin{equation}
 \vec{P}=(\vec{p}-\vec{\acute{p}})^2=\vec{p}^2+ \vec{\acute{p}}^2-2\vec{p}\vec{\acute{p}}.
 \end{equation}
Using the conservation of electron energy with the modified dispersion relations from \cite{15},  where $E_{\pm}^{2}=A^2P^2+\tau P^4\pm 2P\Gamma+m^2$ and setting $h=c=1$, we get:
\begin{equation}
f-\acute{f}=E-E_{0}=\sqrt{A^2P^2+\tau P^4\pm 2P\Gamma+m^2}-m,
\end{equation}
\begin{equation}
(f-\acute{f}+m)^2=A^2P^2+\tau P^4\pm 2P\Gamma+m^2,
\end{equation}
\begin{equation}
P^2=\frac{1}{A^2}[(f-\acute{f})^2+2(f-\acute{f})m-\tau P^4\mp P\Gamma],
\label{Eq4}
\end{equation}
where $f,\acute{f}$ are photon frequency and photon scattering
frequency, $E_{0}$ is the initial energy of the electron in the rest
state, $E$ is the final energy of the electron, $P$ is the momentum
of the electron, $m$ is the mass of the electron,
$\tau=k_{2}L_{p}^2$ and $\Gamma=\frac{k_{3}L_{p}}{2L^{2}}$ are two
positive parameters dependent on helicity propagation of order 1,
where $k_{1}$, $k_{2}$ and $k_{3}$ are unknown dimensional
parameters and $A=1+\frac{k_{1}L_{p}}{L}$ is a Lorentz invariance
violation parameter interpreted as a maximum electron velocity.
$L_{p}$ is the Planck length, and $L$ is a new weave scale. Using
the relation for the Compton effect
($\acute{\lambda}-\lambda=\frac{h}{mc}(1-cos\theta)$), it can be
written \cite{15}:
\begin{equation}
P^2=(f-\acute{f})^2+2f\acute{f}(1-cos\theta).
\label{Eq5}
\end{equation}
Combining  Eqs. \ref{Eq4} and \ref{Eq5}, we get:
\begin{equation}
2f\acute{f}(1-cos\theta)=\frac{1}{A^2}[2(f-\acute{f})m-\tau P^4\mp P\Gamma]+(\frac{1}{A^2}-1)(f-\acute{f})^2.
\label{Eq6}
\end{equation}
The photon scattering frequency shift can be derived from Eq.\ref{Eq6}. Different modified dispersion relations have been proposed, but a modified dispersion relation proposed in Ref. \cite{23,30} is:
\begin{equation}
E^2=P^2+m^2+\frac{|P|^{2+n}}{M^{n}},
\label{Eq7}
\end{equation}
where $M^n$ is a characteristic scale of Lorentz invariance
violation. We can proceed also using relation \ref{Eq7} for modified
dispersion relation. Therefore, a similar above calculation leads to
the following result
\begin{equation}
P^2=(f-\acute{f})^2+2m(f-\acute{f})-\frac{|P|^{2+n}}{M^n}.
\end{equation}
Now, the generalized modified Compton effect of quantum gravity is then:
\begin{equation}
\acute{\lambda}-\lambda=\frac{1}{m}(1-cos\theta)+\frac{\lambda\acute{\lambda}|P|^{2+n}}{2mM^{n}}.
\label{Eq9}
\end{equation}
Rewriting Eq. \ref{Eq9} in the MKS system of units (with order $n=2$), gives:
\begin{equation}
\acute{\lambda}=\frac{\lambda+\frac{h}{mc}(1-cos\theta)}{1-\frac{\lambda|Pc|^{2+n}}{2h(Mc^2)^nmc^3}},
\label{Eq10}
\end{equation}

\begin{equation}
\acute{\lambda}-\lambda=\frac{h}{mc}(1-cos\theta)+\frac{\lambda\acute{\lambda}|Pc|^{4}}{2h(Mc^2)^2mc^3}.
\end{equation}
In this last equation, the frequency shift caused by the LQG effect depends on the frequency of the incident photon.
On the other hand, since the term $\frac{\lambda|Pc|^{2+n}}{2h(Mc^2)^nmc^3}$ is  positive and small, the entire relationship represents a decrease in frequency shift (increase in wavelength shift) \cite{15}.
\section{CMB anisotropies at the last scattering surface}
\label{Sec3}
With the increased accuracy of observational satellites, including the Planck satellite, significant temperature and polarization anisotropies in the CMB radiation have been detected \cite{28,29}.
Although many components of the standard cosmological model ($\Lambda CDM$) are not well understood in terms of fundamental physics,
one  solution to this challenge lies the CMB fluctuations and the high measurement accuracy of observational satellites\cite{31}.\\
The CMB is the remnant of the Big Bang's thermal effects, now cooled down to about $2.725K$ and exhibiting a perfect blackbody spectrum \cite{31}.
CMB temperature anisotropies represent curvature perturbations due to the last scattering surface focused on our current location and spatial fluctuations in the energy density of the CMB.
These perturbations in the CMB temperature anisotropy in direction $\hat{n}$ and time $t_{0}$ are generally expressed by the following equation \cite{32}:
\begin{equation}
\theta(\hat{n})=\theta_{0}+\Psi-\hat{n}.v_{b}+\int_{t_{*}}^{t_{0}}(\dot{\Psi}+\dot{\Phi})dt,
\end{equation}
where $\theta(\hat{n})$ is fractional anisotropy, $\theta_{0}$
represent fractional fluctuations in CMB temperature at the last
scattering surface (time $t_{*}$), $v_{b}$ is the baryon peculiar
velocity, $\Psi$, and $\Phi$ are gravitational potentials (in
general relativity, when $\Psi=\Phi$  the non-relativistic matter is
dominant). At the last scattering surface on larger scales (such as
the Hubble radius), only gravity dominates, while on smaller scales,
the acoustic physics of the primordial plasma and photon diffusion
are
dominated \cite{33}.\\
The main focus of observational CMB research over the past decades has been to obtain accurate estimates of the angular power spectrum of the CMB and relate these values to theoretical models. Research programs like COBE, WMAP, and Planck, along with ground-based telescopes, and the advancements in observational satellite programs, including the latest data from the Planck satellite \cite{28,29}, the parameters of the standard cosmological model have been calculated with increased accuracy. In addition to CMB temperature anisotropy, CMB polarization anisotropy has also been investigated \cite{37}.\\
Polarization is significant in early scattering times because it
enables the growth of a quadrupole anisotropy ($l=2$) around
recombination, while scattering is infrequent until the universe is
re-ionized after the recombination epoch \cite{38}. The linear
polarization expected from the curvature perturbation has an $r.m.s$
of $5\mu$K and is described by two Stokes parameters $Q$, $U$
\cite{31}. Although the Stokes parameters offer a local definition
of polarization, their coordinate dependence is inconvenient for
cosmological interpretation. Instead, linear polarization is
described using the $E$ and $B$ fields \cite{39}.  The Stokes
parameters consist of an orthonormal basis of rank $2$ symmetric and
a trace-free tensor, which can be expressed in terms of the second
derivatives of $E$ and $B$.  Generally,  the Stokes parameters in
Cartesian coordinates, ignoring the sky's curvature, are described
in terms of $E$ and $B$ fields as:
\begin{equation}
\binom{Q\ \ \ \ \ U}{U\ \ \ -Q}\propto (\partial_{i}\partial{j}-\frac{1}{2}\delta_{ij}\nabla^2)E+\epsilon_{k}(_{i}\partial_{j})\partial_{k}B.
\end{equation}
This relation is similar to decomposing a vector field into a gradient part ($E$) and a divergence-free curl part ($B$). E-modes are scalars under parity, while B-modes are pseudo-scalars \cite{31}.  In the absence of parity-violation physics, the two fields should not be correlated, leaving three non-zero polarization power spectra: the $C_{B_{l}}$ B-mode power spectrum, $C_{E_{l}}$ E-mode power spectrum, $C_{C_{l}}$ correlation power spectrum between E-mode and temperature anisotropy; of course, a CMB temperature anisotropy power spectrum, $C_{T_{l}}$, is also presented. Several important points regarding CMB polarization anisotropy include \cite{31}:
\begin{itemize}
    \item[1.]  Polarization is a small signal.
\item[2.]E-mode polarization peaks at scales smaller than temperature because it relies on diffusion in small-scale modes for its generation.
\item[3.]The acoustic peaks in $C_{E_{l}}$ are significant because the temperature quadrupole mainly derives from the bulk plasma velocity, which vanishes when the density is at an extremum.
\item[4.]There is a bump in large-scale polarization caused by re-scattering once the universe re-ionizes.
\item[5.]Symmetrically, B-mode polarizations are not generated by curvature perturbations except through second-order processes like gravitational lensing. This characteristic makes B-mode polarizations a potential probe for gravitational waves.
\end{itemize}
Gravitational waves with wavelength  smaller than the Hubble radius are damped away due to universe's expansion. In this case, CMB is the best option for detecting gravitational waves, as, it is sensitive to early times after the last scattering surface and large scales \cite{40}. Gravitational waves induce CMB temperature anisotropies due to anisotropy expansion effects along the line of sight \cite{41}.
\section{Model Description}
\label{Sec4}
In this section, we present a model that investigates the temperature and E-mode polarization of the CMB anisotropies based on generalized Compton scattering relations within the LQG framework. The first subsection analyzes CMB temperature anisotropy using generalized Compton scattering relations, while the second subsection proposes the anisotropy of the correlation function between temperature and E-mode polarization based on these relations.

\subsection{CMB temperature anisotropy based on the generalized Compton scattering relation}
To study the temporal evolution of CMB anisotropies, the Boltzmann equations are particularly useful, especially when considering the effects of Compton scattering and collision terms. The Boltzmann equation in the synchronous gauge for temperature anisotropy is defined as \cite{42,43}:
\begin{equation}
\dot{\Delta}_{T}^{(S)}+ik\mu\Delta_{T}^{(S)}=-\frac{1}{6}\dot{h}-\frac{1}{6}(\dot{h}+6\dot{\eta})P_{2}(\mu)+\dot{\kappa}\left[-\Delta_{T}^{(S)}+\Delta_{T_{0}}^{(S)}+i\mu v_{b}+\frac{1}{2}P_{2}(\mu)\Pi\right],
\label{Eq13}
\end{equation}
where the derivatives are taken with respect to the conformal
time $\tau$. $\Delta_{T}^{(S)}(\tau,k,\mu)=T^{(S)}$ is the temperature anisotropy of the CMB and the superscript $S$ represents the primordial scalar perturbation in the Fourier modes specified by the wavevector $\mu=\hat{n}.\vec{k}=cos\theta$. Here, $\hat{n}$ represents the direction of photons, $\vec{k}$ is the wavevector, $\theta$ represents the angle between the CMB photon direction $\hat{n}$ and the wavevector $\vec{k}$. $\tau$ is conformal time. $P_{2}(\mu)$ represents Legendre polynomial of order-$2$ ($P_{2}(\mu)=\frac{1}{2}(3\mu^{2}-1)$). $\dot{\kappa}$ is the differential optical depth for Compton scattering, given by:
\begin{equation}
\dot{\kappa}=a(\tau)n_{e}x_{e}\sigma_{T},
\label{Eq14}
\end{equation}
where $a(\tau)$ is the scale factor in terms of conformal time, $n_{e}$ is the number density of free electron, $x_{e}$ is the ionization fraction, and $\sigma_{T}$ is the Thomson cross section. By integrating over conformal time from Eq.\ref{Eq14}, the total optical depth due to the Thomson scattering is obtained as:
\begin{equation}
\kappa(\tau)=n_{e}x_{e}\sigma_{T}\int_{\tau_{r}}^{\tau_{0}}a(\tau)d\tau,
\end{equation}
where $\tau_{0}$ is the present time and $\tau_{r}$ corresponds to
the conformal time of the recombination epoch. The sources in Eq.
\ref{Eq13} include CMB temperature multipoles defined as
$\Delta_{T}^{(S)}(k,\mu)=\sum_{l}(2l+1)(-i)^{l}\Delta_{l}(k)P_{l}(\mu)$,
where $P_{l}(\mu)$ is the Legendre polynomial of order-$l$. $v_{b}$
is the baryon velocity, $\dot{h}$ and $\dot{\eta}$ are the
additional source terms due to the metric perturbation. $\Pi$ is the
source term due to Compton scattering, defined as  \cite{42,43}:
\begin{equation}
\Pi=\Delta_{T2}^{(S)}+ \Delta_{P2}^{(S)}+ \Delta_{P0}^{(S)},
\end{equation}
where $\Delta_{P0}^{(S)}$ and $\Delta_{P2}^{(S)}$ are the source terms due to polarization anisotropy of order $0$ and $2$, respectively.\\
Now, Eq.\ref{Eq13} is modified based on the generalized Compton scattering (Eq.\ref{Eq10} with order $n=2$) of the CMB photon interacting with a resting electron and is contributed a term to Eq.\ref{Eq13} as the generalized Compton scattering effects:
\begin{eqnarray}
\label{Eq4.5}
&&\dot{\Delta}_{T}^{(S)}+ik\mu\Delta_{T}^{(S)}=-\frac{1}{6}\dot{h}-\frac{1}{6}(\dot{h}+6\dot{\eta})P_{2}(\mu)+\dot{\kappa}[-\Delta_{T}^{(S)}+\Delta_{T_{0}}^{(S)}+i\mu v_{b}+\frac{1}{2}P_{2}(\mu)\Pi]\\\nonumber
&&\qquad\qquad\qquad+[i\kappa^{*}\frac{2}{3}\Delta_{T2}^{(S)}(1-\mu^2)+\frac{i+1}{3}(1-\mu^2)\kappa^{*}\Delta_{T1}(\hat{n})]M^{n=2},
\end{eqnarray}
where $\Delta_{T1}(\hat{n})$ is the source due to dipole asymmetry resulting from CMB temperature anisotropy, defined as:
\begin{equation}
    \Delta_{T1}(\hat{n}) =\overline{\Delta_{T1}}(\hat{n})(1+A \hat{p}.\hat{n}).
\end{equation}
$\overline{\Delta_{T1}}(\hat{n})$ is the isotropic part of the
temperature fluctuations, $\hat{n}$ and $\hat{p}$ are the line of
sight (direction of the observation) and the preferred direction
respectively, and $A$ is the amplitude of the asymmetry.
 $M^{n=2}$ from Eq.\ref{Eq10} and $\lambda\propto\frac{1}{f}$ or $(\lambda=\frac{V}{f})$ is rewritten as:
\begin{equation}
M^2=\frac{V^2P^4}{2hc^2(mcV(f-f^{\prime})-hff^{\prime}(1-cos\theta))}
\end{equation}
and $\kappa^{*}$ is the optical depth associated with the generalized Compton scattering,  defined as \cite{44}:
\begin{equation}
\kappa^{*}=\frac{3}{2}\frac{m_{e}v_{e}}{k^{0}}\sigma_{T}\delta_{L}n_{e},
\end{equation}
where $k^0 = 2.7$ Kelvin,  $m_{e}$ electron mass, $v_{e}\sim\frac{1}{\sqrt{1+z}}10^{-3}$ electron bulk flow velocity, $\delta_{L}$ left-handed cosmic electrons density of azimuthal symmetry (and $\delta_{R}$ right-handed cosmic electrons density of azimuthal symmetry) which is assumed in all calculations and this symmetry: $\delta_{L}=\delta_{R}$. For complete solving Eq.\ref{Eq4.5}, we first assume the anisotropies until the present epoch and integrate from all over the Fourier modes:
\begin{equation}
T^{(S)}(\hat{n})=\int d^3k\xi(k)\Delta_{T}^{(S)}(\tau_{0},k,\mu),
\end{equation}
where $\xi(k)$ is a random variable that represents the initial amplitude of the mode. To obtain the power spectrum as an indicator for CMB anisotropies from Eq.\ref{Eq4.5}, we take the integral along the line of sight \cite{41}:
\begin{equation}
\Delta_{T}^{(S)}(\tau_{0},k,\mu)=\int_{0}^{\tau_{0}}d\tau e^{ix\mu}S_T^{S}(k,\tau),
\label{Eq20}
\end{equation}
that
\begin{equation}
S_T^{S}(k,\tau) =\underbrace{ g\left(\Delta_{T0}+2\dot{\alpha}+\frac{\dot{v_{b}}}{k}+\frac{\Pi}{4}+\frac{3\ddot{\Pi}}{4k^{2}}\right)+e^{-\kappa}(\dot{\eta}+\ddot{\alpha})+\dot{g}\left(\alpha+\frac{v_{b}}{k}+\frac{3\dot{\Pi}}{4k^{2}}\right)+\frac{3\ddot{g}\Pi}{4k^2}}_{S_{T1}^{S}}+\nonumber
\end{equation}
\begin{equation}
\underbrace{\left(i\frac{2}{3}\Delta_{T2}^{(S)}(1-\mu^2)+ \frac{i+1}{3}(1-\mu^2)\Delta_{T1}(\hat{n})\right)\kappa^{*}M^2}_{S_{T2}^{S}}.\label{Eq21}
\end{equation}
$S_{T1}^{S}$ of Eq.\ref{Eq21} is the power spectrum resulting from the standard cosmological scenario, and $S_{T2}^{S}$ is the power spectrum resulting from the generalized Compton scattering effect.  $x=k(\tau_{0}-\tau)$ and $\alpha=\frac{\dot{h}+6\dot{\eta}}{2k^2}$. We have introduced the visibility function $g(\tau) = \dot{\kappa}\exp(-\kappa)$. \\
Now, the power spectrum of temperature anisotropy as a rotational invariance quantity is defined as $C_{T_{l}}=\frac{1}{2l+1}\sum_{l}|a_{lm}|^{2}$,
\begin{equation}
    C_{Tl} = \frac{1}{2l+1}\sum_{l}\langle a^\star_{T,lm}a_{T,lm}\rangle,
\end{equation}
in terms of which,
\begin{equation}
\langle a^\star_{T,lm}a_{T,lm}\rangle = C_{Tl}\delta_{l'l}\delta_{m'm}.
\end{equation}
 Here, $a_{lm}$ represents the amplitude of CMB temperature fluctuations in the presence of scalar perturbation, given by \cite{45}:
\begin{equation}
a_{T,lm}=\int d\Omega Y_{lm}^{*}(\hat{n})T^{(S)}(\hat{n}),
\end{equation}
where $\Delta_{T}$ is the CMB temperature anisotropy obtained from Eqs.\ref{Eq20}, \ref{Eq21} and $Y_{im}^{*}(\hat{n})$ is the spherical harmonic function. Finally, the CMB temperature anisotropy power spectrum obtained from Eqs.\ref{Eq20} and \ref{Eq21} is equal to:
\begin{eqnarray}
&&C_{T_{l}}=\frac{1}{2l+1}\int d^3k\phi_{k}\sum_{l}|\int_{0}^{\tau_{0}}d\Omega Y_{lm}^{*}(\hat{n})\Delta_{T}^{(S)}(\tau_{0},k,\mu)|^{2}\nonumber\\
&&=16\pi^2\int k^2\phi_{k}dk\left[\left(S_{T1}^{(S)}(\tau_{0},k,\mu)j_{l}(x)\right)^2+\left(S_{T2}^{(S)}(\tau_{0},k,\mu)\frac{j_{l}(x)}{x^2}\right)^2\right],
\end{eqnarray}
where $\phi_{k}$ is the initial power spectrum ($\phi_{k}=pn_{s}-1$ where $p$ is the photon momentum), $j_{l}(x)$ is the spherical Bessel function of order-$l$ and assuming the framework $k\parallel\hat{z}$, $\int d\Omega Y_{lm}^{*}(\hat{n})e^{ix\mu}= \sqrt{4\pi(2l+1)}i^{l}j_{l}(x)\delta_{m0}$.

\subsection{Anisotropy of the correlation function between temperature and E-mode polarization based on the generalized Compton relation}
The correlation function between temperature and E-mode polarization
is established, while the correlation between (temperature and
B-mode) or (E-mode and B-mode) vanishes because B-mode has the
opposite parity to $T$ and $E$ \cite{46}. This CMB correlation
function is defined as follows \cite{45}:
\begin{equation}
C(\theta)=\sum_{l}\frac{2l+1}{4\pi}C_{C_{l}}^{(S)}P_{l}cos(\theta),
\end{equation}
where $P_{l}$ is the Legendre polynomial of order $l$ and $\theta$
is an angle scale. The angular power spectra are in terms of
multipole $l$ values, and the correlation function is in terms of
the angular scale $\theta$. The relationship
$\theta\propto\frac{1}{l}$ indicates that large multipole values
correspond to small angular scales \cite{47}. Generally, the
correlation power spectrum between $T$ and $E$ is defined as
\cite{46}:
\begin{equation}
C_{C_{l}}=\frac{1}{2l+1}\sum_{m}<a_{T,lm}^{*},a_{E,lm}>.
\end{equation}
In the presence of scalar perturbation for the correlation function,
we have:
\begin{equation}
\Delta_{C_{l}}^{(S)}=\int_{0}^{\tau_{0}}\dot{\kappa}e^{ix\mu-\kappa}\Pi_{\mu}^{2}(1-\mu^2)^2\kappa^{*}\Delta_{T_{l}}^{(S)}(\tau_{0},k,\mu)\Delta_{E_{l}}^{(S)}(\tau_{0},k,\mu).
\end{equation}
Finally,  the power spectrum of the correlation function in the presence of scalar perturbation is written as:
\begin{equation}
C_{C_{l}}^{(S)}=16\pi^2\int k^2dk\phi_{k}\Delta_{T_{l}}^{(S)}(\tau_{0},k,\mu)\Delta_{E_{l}}^{(S)}(\tau_{0},k,\mu).
\end{equation}
\section{Modified-$\Lambda$CDM verse data}
\label{Sec5}
\begin{figure}
    \begin{center}
        \includegraphics[height=4cm,width=7cm]{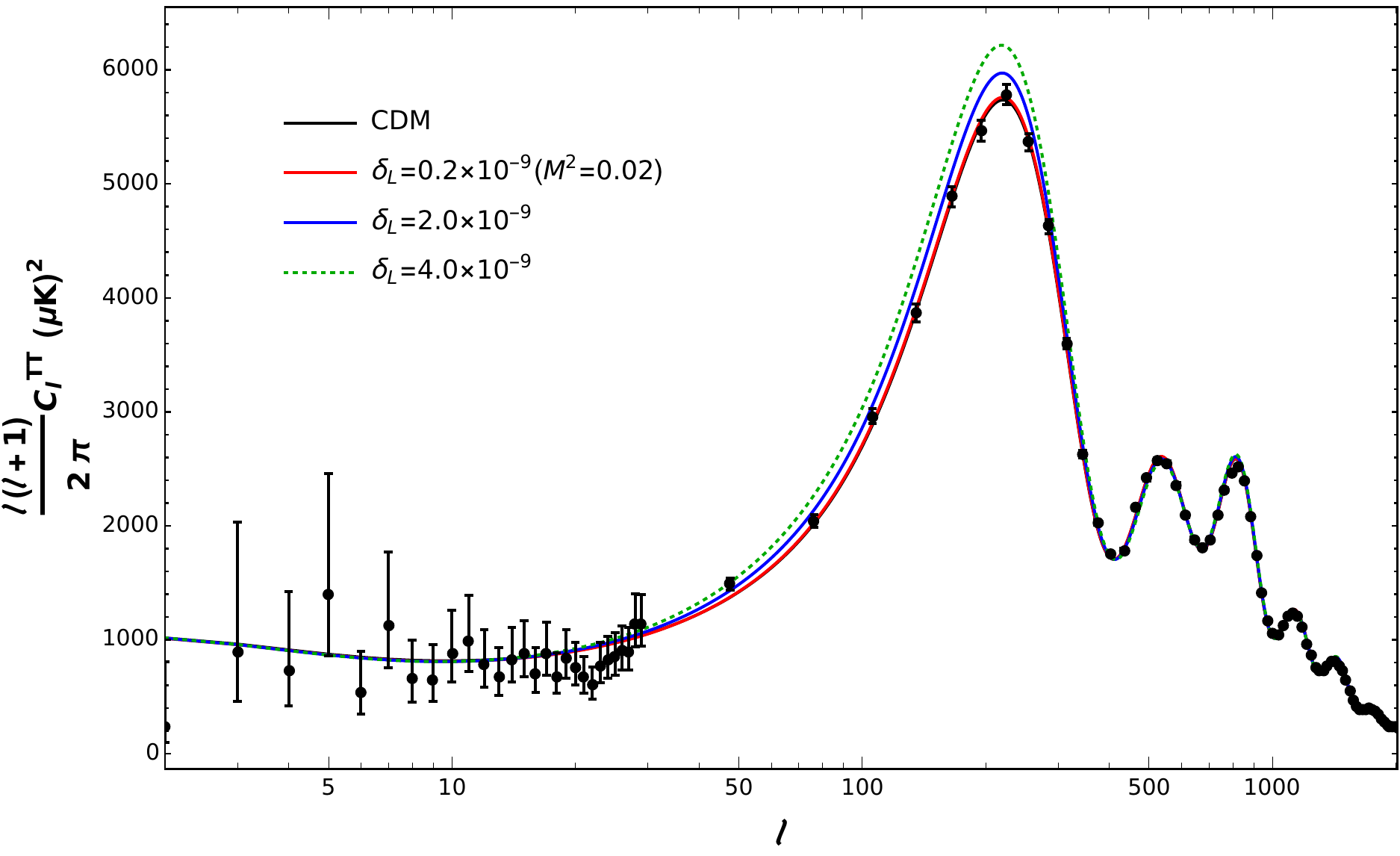}
        \includegraphics[height=4cm,width=7cm]{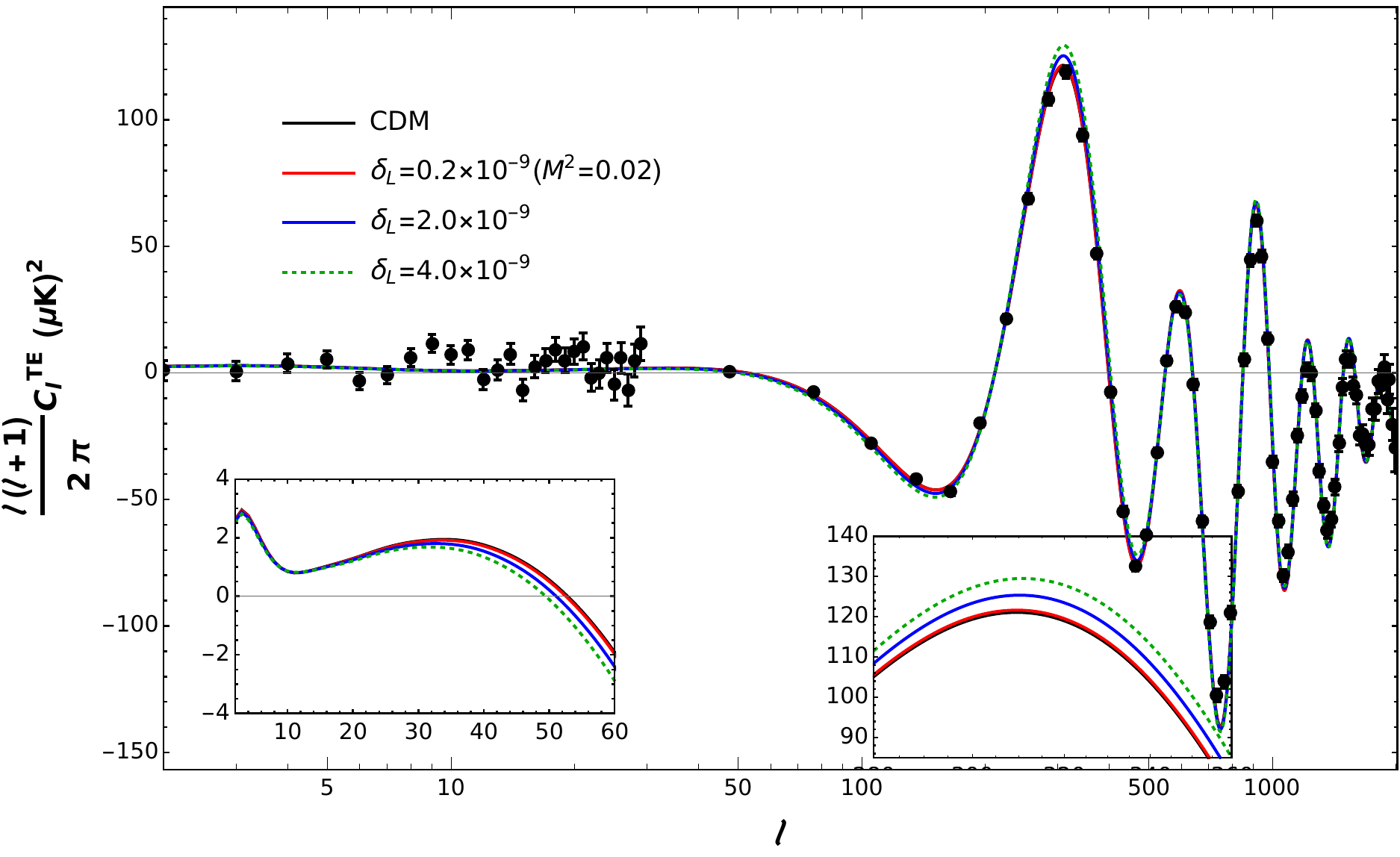}
        \includegraphics[height=4cm,width=7cm]{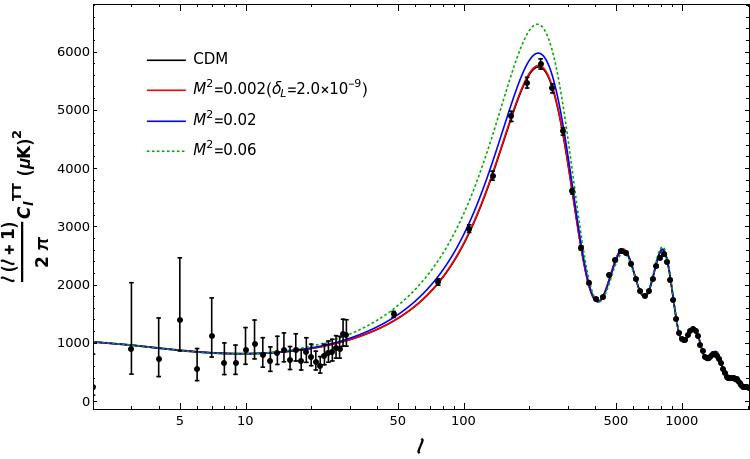}
        \includegraphics[height=4cm,width=7cm]{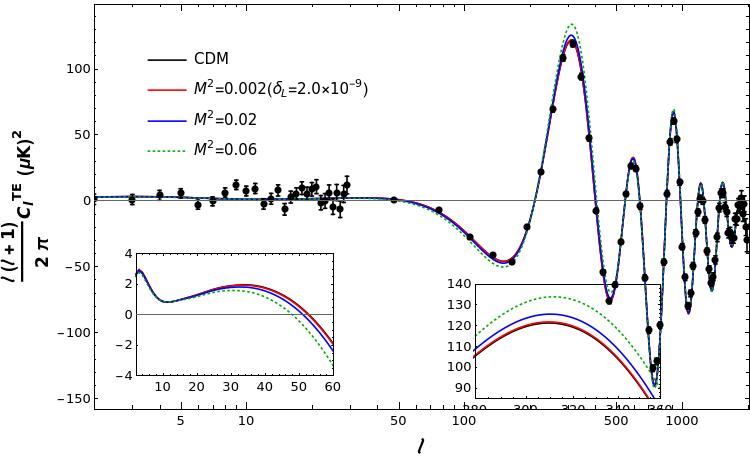}
        \caption{Left panels: Temperature anisotropies in the CMB. Right panels: Anisotropy of the correlation function between temperature and E-mode polarization in the CMB.
        }
        \label{fig-1}
    \end{center}
\end{figure}

To check the anisotropy observed in CMB
radiation based on a modified-generalized Compton scattering with loop quantum gravity
approach, we implement the related equations in the publicly available numerical code \texttt{CLASS}\footnote{\label{myfootnote0}\url{https://github.com/lesgourg/class_public}}(the Cosmic Linear Anisotropy Solving System)~\citep{Lesgourgues:2011rh}. In module perturbation.c of \texttt{CLASS}, the temperature source function is divided into three distinct parts:
\begin{equation}
    S_T =S_T^0+\frac{d}{d\tau}S_T^1+ \frac{d}{d\tau}\dot{S}_T^2,
\end{equation}
where the dot indicates that the code evaluates the analytic expression for the derivative, while the derivatives  denoted by $\frac{d}{d\tau}$ are estimated numerically. Therefore, it is useful to provide the explicit expressions for all three terms in the Newtonian gauge as
\begin{eqnarray}
    &&S_T^0 = g(\frac{1}{4}\delta_\gamma+\phi)+2e^{-\kappa}\dot{\phi}+\frac{1}{k^2}g\dot{\kappa}\theta_b+e^{-\kappa} \ddot{\kappa}\theta_b+e^{-\kappa}\dot{\kappa}\dot{\theta}_b,\\ \nonumber
    &&S_T^1 = e^{-\kappa} k(\psi-\phi),\\ \nonumber
    &&S_T^2 =\frac{1}{8} g \Pi.
\end{eqnarray}
We add the modified terms to $S_T^0$ and $S_T^2$.\\
In addition to the six parameters of the standard $\Lambda$CDM
model, namely baryon density $\Omega_b h^2$, dark matter density
$\Omega_{DM} h^2$, the ratio between the acoustic scale and the
angular diameter distance at decoupling $\theta_s$, optical depth of
reionization $\tau_{reio}$, the amplitude $A_s$, and the spectral
index $n_s$, two additional parameters $M^2$ ad $\delta_L$, are
included for the applied modifications. The value of $\delta_L$, the density caused by cosmic electrons, often referred to as the optical depth to reionization ($\tau$), is reported to be of the order of $0.054 \pm 0.007$ in the Planck 2018 data \cite{28}.  The optical depth to reionization is a crucial parameter in cosmology, as it quantifies the integrated number of free electrons along the line of sight back to the epoch of reionization, affecting the amplitude of the large-scale E-mode polarization in the cosmic microwave background. The value of $\delta_L$ obtained, of order $10^{-9}$, corresponds to the reported value of the optical depth to reionization by the Planck 2018 release.\\
 Another free parameter in the study is
$M^2$. It is crucial to accurately estimate the prior values of
the $M^2$ parameter due to its significant impact on the results.
Given the absence of the established prior information for the $M^2$
parameter, an exploratory approach was adopted to test various
potential values. We considered different orders of magnitude for $M^2$, as shown in the bottom of Figure~\ref{fig-1}, and finally considered its prior in the interval $[0, 0.1]$. In the panels of Figure~\ref{fig-1}, we observe an increase in the amplitude of the lower multipoles in the temperature and the correlation function power spectrum until the first peak. The integrated Sachs-Wolfe (ISW) effect is important on such scales, and it is possible that the modified-generalized Compton scattering affects this signal.\\
\begin{figure}
    \begin{center}
        \includegraphics[height=4cm,width=7cm]{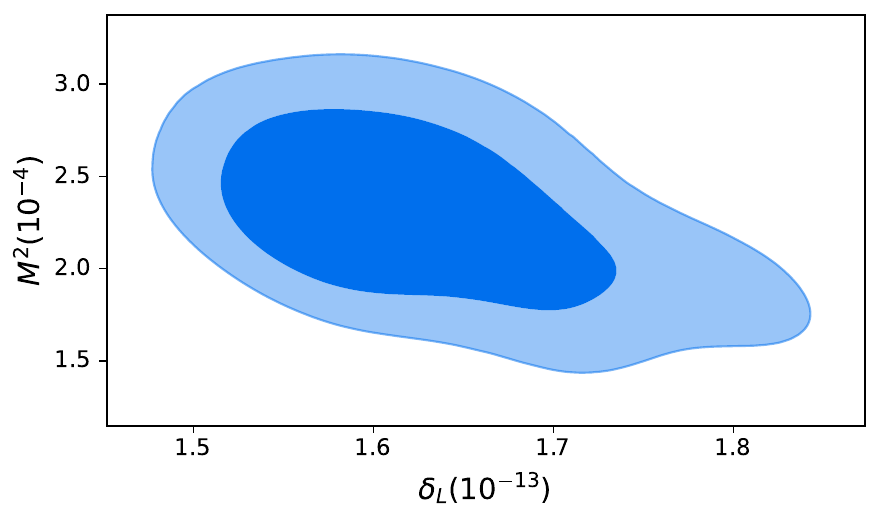}

        \caption{1D likelihoods and 2D contours in 68\% and 95\% CL marginalized joint regions for chosen free parameters: $\delta_L$ and $M^2$.
        }
        \label{fig-2}
    \end{center}
\end{figure}
\begin{figure}
    \begin{center}
        \includegraphics[height=4cm,width=7cm]{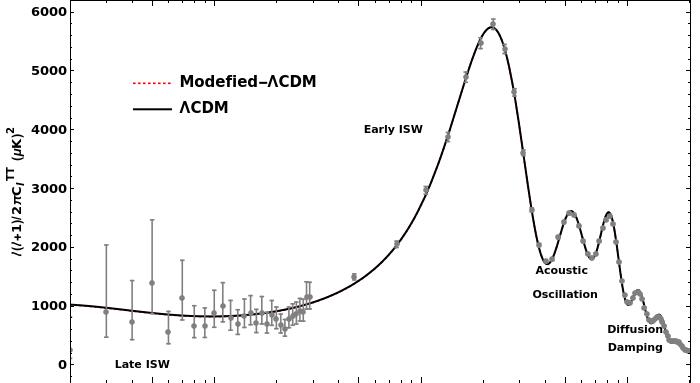}
        \includegraphics[height=4cm,width=7cm]{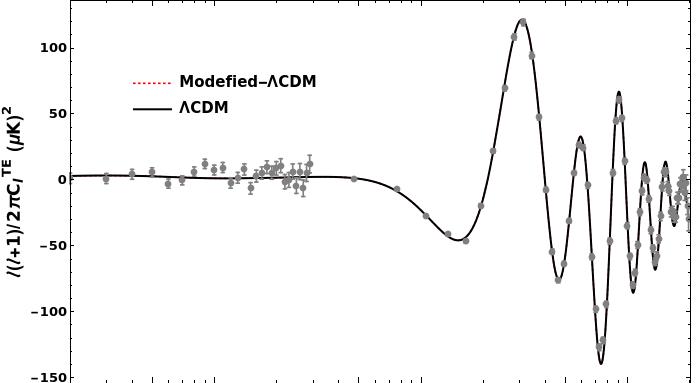}
            \includegraphics[height=2cm,width=7cm]{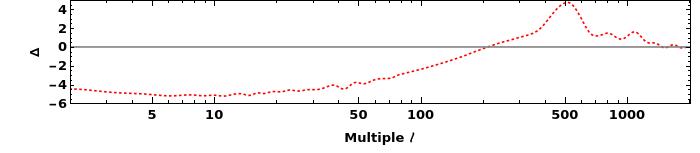}
        \includegraphics[height=2cm,width=7cm]{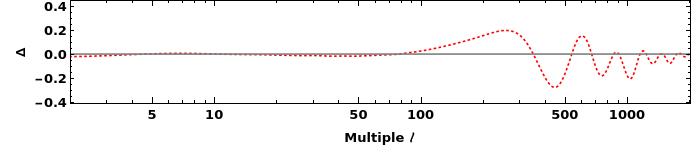}

        \caption{The temperature and TE
            cross correlation CMB angular power spectra using the best fit values of
            cosmological parameters. The red curve in the plots below indicates the deviation from the standard model.  }
        \label{fig-3}
    \end{center}
\end{figure}
To obtain the exact values of $\delta_L$ and $M^2$, we use the code
\texttt{MONTEPYTHON-v3}\footnote{\label{myfootnote}\url{https://github.com/baudren/montepython_public}}~\citep{Audren:2012wb,Brinckmann:2018cvx}
to perform a Monte Carlo Markov Chain (MCMC) analysis with a
Metropolis-Hasting algorithm. We use the CMB temperature and
polarization auto- and cross-correlation measurements of the most
recent Planck 2018 legacy release, including the full temperature
power spectrum at multipoles 2 $\leq$ l $\leq$ 2500 and the
polarization power spectra in the range 2 $\leq$ l $\leq$ 29. We
also include information on the gravitational lensing power spectrum
estimated from the CMB trispectrum analysis \cite{29}. We obtain the
following values using all CMB and CMB lensing information:
\begin{eqnarray}
&&  M^2=2.28\pm0.34(10^{-4}), \nonumber\\
&&  \delta_L=1.63\pm0.08(10^{-13}).\nonumber
\end{eqnarray}
We show posterior distributions ($1\sigma$ and $2\sigma$ intervals) of these parameters in Figure~\ref{fig-2}. Given the small values obtained for these parameters, we do not expect to observe any deviation in anisotropy observed in CMB radiation compared to the standard model, as seen, for example, in Figure~\ref{fig-3} for temperature anisotropies and TE
cross correlation CMB angular power spectra.\\
To select the most compatible model with the observational data and
assess the quality of the fit, we use the least squares method,
commonly applied in cosmology, known as the total chi-squared,
$\chi^2_{{}_{\rm tot}}$. A model with a smaller $\chi^2_{{}_{\rm
tot}}$ is generally considered a better fit to the
data~\citep{Davari:2021mge}. In our analysis, the $\chi^2_{{}_{\rm
tot}}$ values for the standard model and the Modified-$\Lambda$CDM
model are $1389.47$ and $1388.30$, respectively. However, a lower
$\chi^2_{{}_{\rm tot}}$ does not necessarily indicate the best
model, as models with more parameters can artificially lower
$\chi^2_{{}_{\rm tot}}$. To address this, we employ the Akaike
Information Criterion (AIC), defined as $AIC=\chi^2_{\rm min}+2M$,
where $M$ is the number of free parameters in the model. According
to \cite{50}, $\Delta AIC$ values greater than 2, 5, and 10 indicate
weak, moderate, and strong evidence respectively, against the model
with the higher $AIC$ value. In this study, we consider two
additional parameters compared to the standard model, and since
$|\Delta AIC|=2.18$, the two models fit the cosmological data
equally well.
\section{Conclusion}
\label{Sec6}
In this study, we explored the modified-generalized Compton scattering relations within the framework of loop quantum gravity and analyzed the anisotropies in the cosmic microwave background (CMB) radiation's temperature and polarization, influenced by processes originating from the last scattering surface.

We developed a model to account for temperature anisotropy and the anisotropy in the correlation function between temperature and E-mode polarization. This model integrates the modified-generalized Compton scattering relations and leverages the latest Planck satellite data (2018), incorporating two additional parameters: $\delta_{L}$ (cosmic electron density) and $M^{2}$ (effects of modified-generalized Compton scattering).

Our findings indicate that smaller values of $\delta_{L}$ and $M^{2}$ yield results that closely align with the standard cosmological model, corroborating the effects of modified-generalized Compton scattering with the 2018 Planck data. As depicted in Figure~\ref{fig-1}, increasing the values of $\delta_{L}$ and $M^{2}$ enhances the fluctuation range in the CMB temperature anisotropy power spectrum and the correlation function between temperature and E-mode polarization for multipoles $l<500$ up to the first peak. Conversely, for multipoles $l>500$, variations in these parameters do not significantly impact the results.

We anticipate that future observations with improved precision from new missions or experiments will enable the determination of more accurate values for $\delta_{L}$ and $M^{2}$. Such enhanced accuracy may reveal substantial deviations from the standard model. In future research, we aim to further examine the implications of modified-generalized Compton scattering on both E-mode and B-mode polarization, potentially providing deeper insights into the fundamental structure and evolution of the universe.
\section{Acknowledgments}
Z.D. was supported by  the Korea Institute for Advanced Study Grant No 6G097301.\\
S.D. Sadatian and A. Sabouri appreciate R. Mohammadi for helpful
discussions and useful advices during this study.
\section{Data availability}
No new data were generated or analysed in support of this
research.
\section*{Declarations Conflict of interest}
The authors declare that they have no conflict of interest.

\end{document}